\documentclass[epjCONF,columns]{svjour}
\usepackage{graphics}
\usepackage[varg]{txfonts} % Times fonts
\usepackage[latin1]{inputenc}
\session-title{Hadron Collider Physics (HCP) 2011, Paris, France}
\begin{document}
\title{Soft QCD, minimum bias and diffraction: results from ALICE}
\author{Pietro Antonioli\inst{1}\fnmsep\thanks{\email{pietro.antonioli@bo.infn.it}} for the ALICE Collaboration}
\institute{INFN - Sezione di Bologna, Via Irnerio 46, 40126 Bologna, Italy}
\abstract{We report recent results from the ALICE experiment at the LHC for minimum bias pp collisions.
This overview includes results on inelastic cross section, with analysis of single and double 
diffractive events; the study of hadron production mechanisms, both for inclusive and 
identified particles; Bose-Einstein correlations; and fluctuations in $\langle p_T\rangle$.
}
\maketitle
\section{Introduction}
\label{intro}
While the main focus of the ALICE experiment~\cite{AliceGen} at the LHC is the study of relativistic heavy-ion
collisions, ALICE also has a rich and unique proton-proton physics programme, arising from its design. The use of a moderate magnetic field (B=0.5 T) in the barrel region
coupled with little material close to the interaction point (7\% $X_0$ perpendicular to the beam direction) allows the
study of particle spectra down to very low $p_T$ ($\sim$100 MeV/c). 
In the barrel region ALICE has extensive particle identification
capabilities, for charged hadrons via measurements of dE/dx in the inner silicon detector (ITS) and Time Projection 
Chamber (TPC), in the time-of-flight detector (TOF), and by a devoted transition radiation detector for electrons. 

Results summarized here are primarily from data collected in 2010 pp run at $\sqrt s$=7 TeV, with a total of 300 million events
analyzed, collected with a minimum bias trigger. The trigger requires a hit in the inner silicon detector (SPD) or in
either two scintillator counters arrays (VZERO) positioned at z=3.3 m and 0.9 m from the interaction point,
close to the beam pipe. The minimum bias trigger essentially requires at least one charged particle
anywhere in 8 units of rapidity. Data collected at $\sqrt s$=2.76 TeV in 2011 have been also analyzed, with
a total of 65 million events. Section~\ref{sec:crosssec} present measurements related to inelastic cross section and diffraction,
while section~\ref{sec:prod} reports several measurements studying production mechanisms, taking advantage
of particle identification capabilities of the detector. Section~\ref{sec:hbt} presents general
events characterstics, obtained by means of Bose-Einstein correlations and studies of fluctuations in $\langle p_T\rangle$. Other proton proton soft QCD topics
actively studied in ALICE and not reported here include the study of $\pi ^0$ yield to constrain 
the gluon fragmentation function and  characterization of underlying event observables.  
Results related to heavy flavours in proton proton collisions are presented elsewhere in this conference~\cite{hcpHF}.

\section{Inelastic cross section and diffraction}
\label{sec:crosssec}
\begin{figure}[t!]
\begin{center}
\resizebox{0.75\columnwidth}{!}{\includegraphics{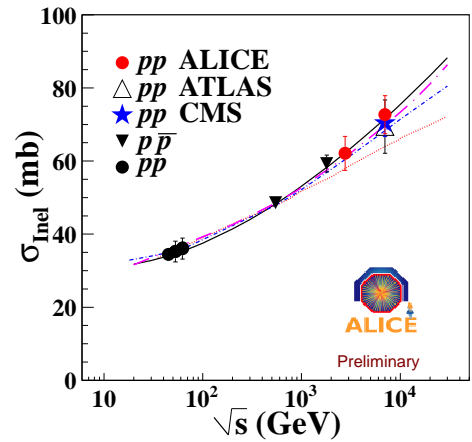} }
\caption{Inelastic cross-section as a function of collision energy. Data are compared with the predictions of~\cite{diffModel} (solid black line), 
~\cite{diffOst} (long dot-dashed pink line),~\cite{diffGot} (short dot-dashed blue line) and~\cite{diffRysk} (dotted red line).}
\label{fig:inel}   
\end{center}
\end{figure}

To obtain the total inelastic cross section for pp collisions, the cross section of a reference
trigger process was measured. This is needed to scale the normalization. The re\-ference cross
section has been obtained by means of the Van der Meer scan method, which was applied
at $\sqrt s$=7 TeV and 2.76 TeV, with the latter scan performed in March 2011. A trigger based on the
VZERO detector was used. The cross section for production of at least one charged 
particle with $p_T >$0.5 GeV/c in the pesudorapidity region $|\eta|<$0.8 is in good
agreement between LHC experiments and results in the measurement of the total
inelastic cross-section shown in Fig.~\ref{fig:inel}, compared with results from ATLAS and CMS.
The TOTEM measurement recently published ~\cite{TOTEM} is also in agreement.

For the analysis of the inelastic cross section ALICE used three subdetectors: the VZERO, the silicon pixel
detector (SPD), and the forward multiplicity detector (FMD), an array of silicon sensors
at large rapidity. With the addition of the FMD the pseudorapidity coverage is -3.7$<\eta<$5.1.
Due to the geometry of the detectors used it was possible to define one and two-arm triggers and
to study the pseudorapidity gap pattern of the tracks obtained from the event vertex and a hit
in SPD, VZERO or FMD. With this technique ALICE measured the relative fraction of single diffractive
(SD) and double diffractive (DD) processes. Since the non-diffracted proton in SD processes is outside detector
acceptance in ALICE, the measurement is model dependent. PYTHIA and PHOJET Monte Carlo generators
were tuned using the model described in~\cite{diffModel} to provide the SD cross-section dependence
on the diffracted mass.

\begin{figure}[t!]
\begin{center}
\resizebox{0.75\columnwidth}{!}{\includegraphics{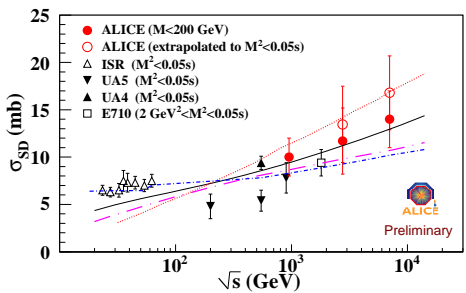} }
\caption{Single diffractive cross-section as a function of collision energy. Predictions are from the same
models as in Fig.~\ref{fig:inel}}.
\label{fig:sd}   
\end{center}
\end{figure}

\begin{figure}[t!]
\begin{center}
\resizebox{0.75\columnwidth}{!}{\includegraphics{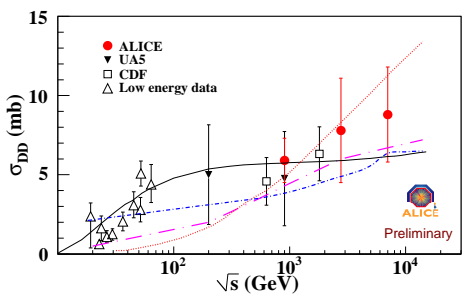} }
\caption{Double diffractive cross-section as a function of collision energy. Predictions are from the same
models as in Fig.~\ref{fig:inel}}.
\label{fig:dd}   
\end{center}
\end{figure}

The result for SD and DD cross section are shown in Fig.~\ref{fig:sd} and ~\ref{fig:dd}, compared with previous
measurements at lower center of mass energies. There is good agreement between ALICE and UA5 at 
$\sqrt s$=900 GeV. Within the errors, the theoretical models broadly describe the data, 
though they all have a difficulty to fit the data simultaneously at lower and higher energies, especially
for DD processes. The ratio of $\sigma _{SD}$ and $\sigma _{DD}$ with respect to the inelastic
cross sections is constant within the uncertainty. Further details can be found in~\cite{pogQM}.

\section{Production mechanisms}
\label{sec:prod}
\subsection{Inclusive production}
\label{sec:inclusive}
\begin{figure}[h!]
\begin{center}
\resizebox{0.75\columnwidth}{!}{\includegraphics{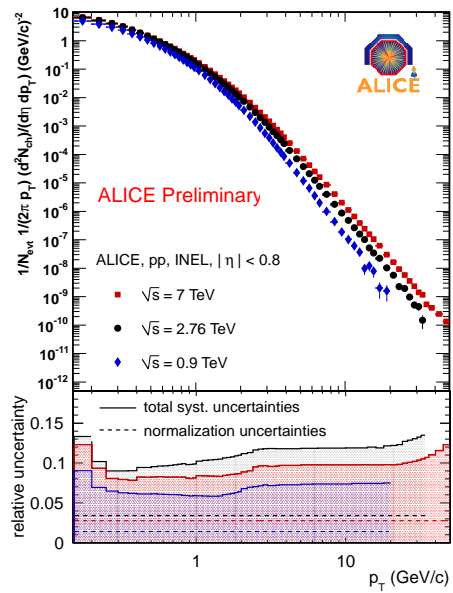} }
\caption{Charged particle transverse momentum spectrum (in the range 0.15$<p_T<$50 GeV/c) measured in pp collisions at $\sqrt{s}$ = 0.9, 2.76 and 7 TeV with 
the total systematic uncertainties as well as the overall normalization uncertainty.}
\label{fig:spe276}   
\end{center}
\end{figure}

Charged multiplicity and inclusive spectra of charged particles have been extensively studied by ALICE,
which has reported results at different collision energies ($\sqrt s$=0.9~\cite{mult900}, 2.36~\cite{mult900},
and 7 TeV~\cite{mult7TeV}. During 2011 a low-energy run at $\sqrt{s}$=2.76 TeV was undertaken to
provide accurate norma\-lization for Pb-Pb data which were collected 
at the same center of mass energy. The observed charged particle
$p_T$ spectrum is shown in Fig.~\ref{fig:spe276}, together with previous measurements. A 
modified Hagedorn function successfully fits the data, with a $p_{T}^{-n}$ power law
at high $p_T$ (above 3 GeV/c). Note the low $p_T$ reach of the detector, which has been used
to find discrepancies with pre-LHC Monte Carlo tunes below 0.5 GeV/c ~\cite{mult900}.

\begin{figure}[h!]
\begin{center}
\resizebox{0.95\columnwidth}{!}{\includegraphics{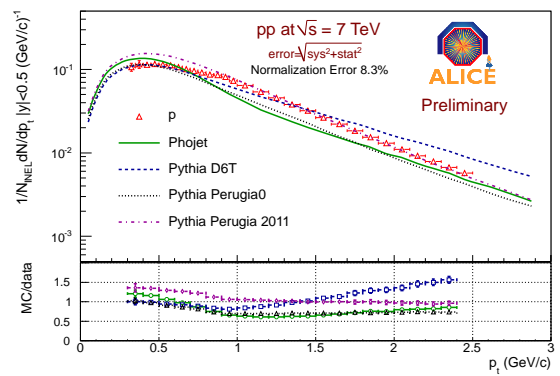} }
\caption{Measured proton spectra compared with pre-LHC and more recent MC tunes.
}
\label{fig:mcspec}   
\end{center}
\end{figure}

\subsection{Identified particle spectra}
\label{sec:partid}
\begin{figure}[h!]
\begin{center}
\resizebox{0.93\columnwidth}{!}{\includegraphics{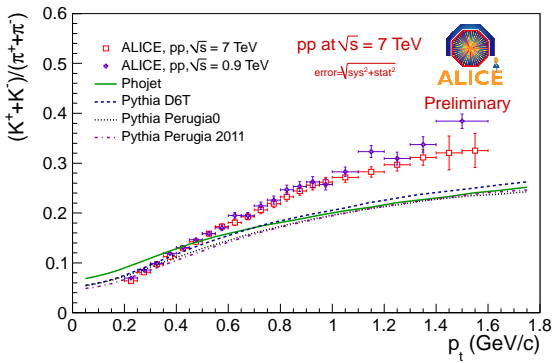} }
\caption{Kaon to pion ratio compared to various MC generators.
}
\label{fig:partrat}   
\end{center}
\end{figure}

The ALICE experiment, using its particle identification capabilities, provided preliminary results 
for identified particles spectra for pions, kaons and protons. 
These are based on individual or combined measurements made by its ITS, TPC and TOF detectors in various $p_{T}$ ranges.
A L\'evy-Tsallis function fits the data and allows to extrapolate the yield down to $p_T$=0.
Comparisons are made with several Monte Carlo codes. PYTHIA tune Perugia 2011
shows good agreements with kaons and overestimates the pion yield, while the proton description is in
good agreement with data above 0.7 GeV as seen in Fig.~\ref{fig:mcspec}. The study of particle ratios (shown in Fig.~\ref{fig:partrat})
challenges Monte Carlo tunes. No discernible energy dependence was obtained for the
particle ratios, when compared with existing measurements made by PHENIX, STAR, E735 and UA5, while
the $\langle p_T\rangle$ showed a modest increase with energy for pions, kaons and protons, consistently with a a linear
expectation from a simple linear scaling.

\subsection{Strange and multi-strange hadrons}
\label{sec:strange}
\begin{figure}[h!]
\begin{center}
\resizebox{0.95\columnwidth}{!}{\includegraphics{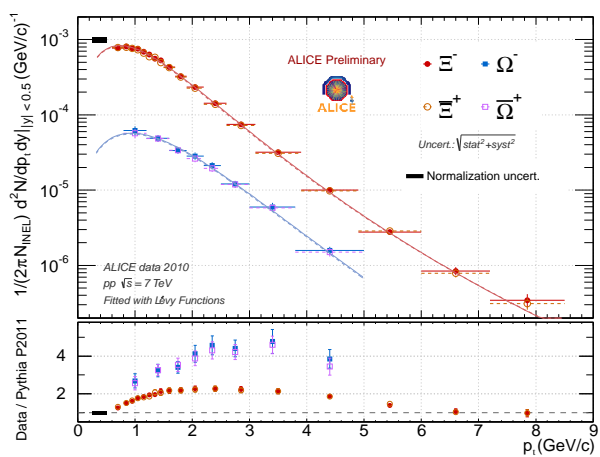} }
\caption{
$\Omega^{\pm}$ and $\Xi^{\pm}$ spectra measured at ALICE, compared with Perugia 2011 tune.}
\label{fig:omegaxi}   
\end{center}
\end{figure}

Particle identification methods were combined with weak decays topological techniques and invariant mass
analyses to identify and study strange and multi-strange hadrons. 
PYTHIA tune Perugia 2011 provides good predictions for kaons, but the agreement is not yet satisfactory
for $\Xi^{\pm}$ and $\Omega^{\pm}$, in particular in the intermediate (1-5 GeV/c) $p_T$ region as seen in 
Fig.~\ref{fig:omegaxi}. The agreement at high $p_T$ (6-9 GeV/c) is good for $\Xi^{\pm}$, 
but further statistics is needed for $\Omega^{\pm}$. 
It should be noted that the Perugia 2011 tune optimizes strangeness 
production using early LHC measurements, including from ALICE.
However, it underestimates the production of strange resonances, as $\Sigma^{*}$ and $K^{*0}$, while various
MC tunes shows better agreement with $\phi$, especially PHOJET. The differences are mainly related to ba\-ryons
productions, suggesting the possible presence of other ha\-dronization mechanisms, like quark coalescence,
which ha\-ve been proposed to explain the enhancement of the yield ratio $\Lambda/K^0_S$ 
reported in heavy ions collisions. 

\section{Bose-Einstein correlations and fluctuations}
\label{sec:hbt}
Bose-Einstein correlations of identical boson pairs
(pions or kaons) have been studied by ALICE. The extraction of Hanbury-Brown Twiss radii, 
and their dependence on va\-rious observables, may shed light on the spatial scale of the emitting source. 
This femtoscopic analysis in pp collisions is able to obtain precise data on 'elementary' systems. 
High-multiplicity pp collisions at the LHC have particles densities comparable
to that measured at RHIC in heavy ions, and a direct comparison of the freeze-out
sizes of systems with very different initial states is therefore possible. 

The correlation function for identical pion pairs is defined
as $C\left({\bf q}\right)=A\left({\bf q}\right)/B\left({\bf q}\right)$, where  $A\left({\bf q}\right)$ is the measured  
two-pion distribution of pair momentum difference ${\bf q}={\bf
p}_2-{\bf p}_1$, and $B\left({\bf q}\right)$ is the equivalent 
distribution obtained by using pairs of particles coming from different
events. A quantitative analysis of the correlation functions allows
the extraction of the sizes of the emitting source (HBT radii) $R^{G}_{long}$, 
$R^{G}_{out}$ and $R^{G}_{side}$, which are oriented along the beam axis, along the pair
transverse momentum $k_T$ and perpendicular to the other two. Full details
are described in~\cite{becAlice}.

\begin{figure}[b!]
\begin{center}
\resizebox{0.90\columnwidth}{!}{\includegraphics{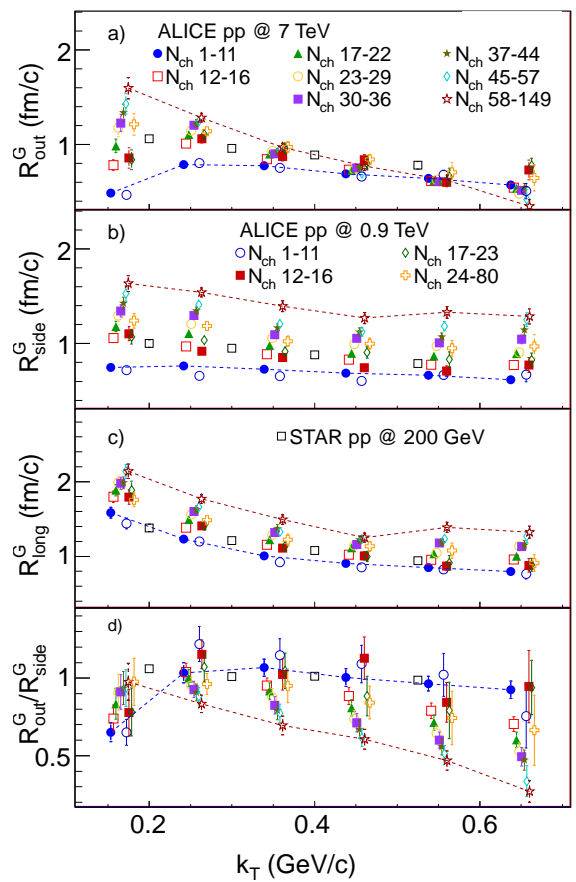} }
\caption{HBT radii dependence on $k_T$ for different multiplicities, with
ALICE data at $\sqrt s$=0.9 TeV and $\sqrt s$=7 TeV. 
}
\label{fig:hbtkt}   
\end{center}
\end{figure}

Fig.~\ref{fig:hbtkt} shows dependence of the HBT radii on the
pair transverse momentum, for various event multiplicities. We observe
$R^{G}_{out}$ and $R^{G}_{side}$ decrease with $k_T$ at large multiplicities,
while $R^{G}_{long}$ falls at all multiplicities. In heavy ions collisions the decrease
with $k_T$ is a signature of collective motion. While the observed behaviour
in pp is qualitatively similar, the difference seen in particular for the transverse
radii shows that a radial flow interpretation cannot be ea\-sily inferred.

\begin{figure}[t!]
\begin{center}
\resizebox{0.90\columnwidth}{!}{\includegraphics{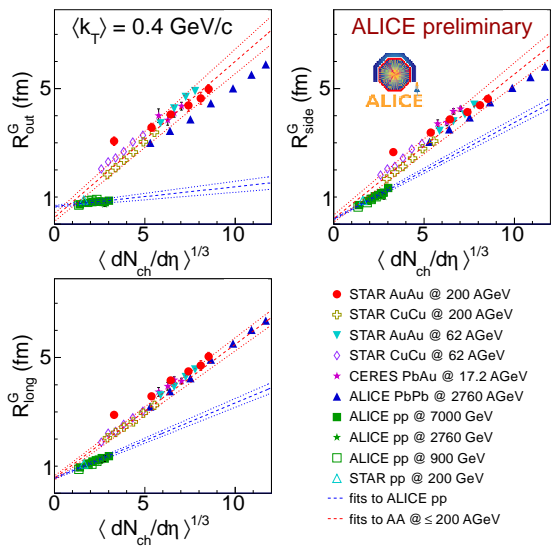} }
\caption{HBT radii as a function of charged pseudorapidity density. Results
from pp and heavy ions collisions are compared.
}
\label{fig:hbtdn}   
\end{center}
\end{figure}

A linear dependence with $\langle dN/d\eta \rangle^{1/3}$ is expected for the HBT radii. 
Fig.~\ref{fig:hbtdn} shows the direct comparison between elementary and
'compound' systems: different slopes and offsets for pp and heavy ions are observed
and the size of the pp emitting source is $\approx$ 1 fm. The linear fits
show qua\-litative agreement with hydrodynamical models, though the 
trends inferred by other lower energies heavy ions data are not in agreement with $R^{G}_{out}$.

\begin{figure}[t!]
\begin{center}
\resizebox{0.75\columnwidth}{!}{\includegraphics{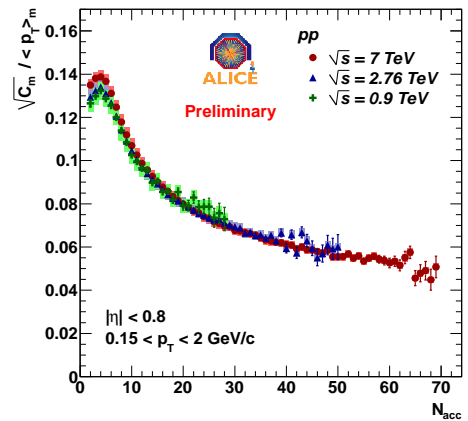} }
\caption{Relative fluctuations in ALICE at different center of mass energy as a function
of accepted tracks multiplicity
}
\label{fig:fluc-norm}   
\end{center}
\end{figure}

\begin{figure}[ht]
\begin{center}
\resizebox{0.75\columnwidth}{!}{\includegraphics{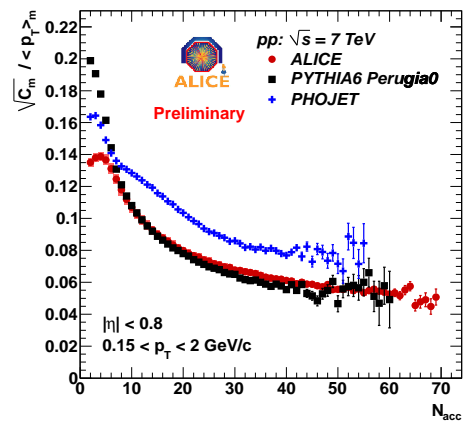} }
\caption{Relative fluctuations: comparison of ALICE pp data with MC generators.
}
\label{fig:fluc-norm-mc}   
\end{center}
\end{figure}

Global event characterization was also studied mea\-suring fluctuations
of $\langle p_T\rangle$ in pp collisions. This observable is expected
to be dominated by resonance decays, Bose-Einstein
correlations and mini-jets. We study fluctuations of $\langle p_T\rangle$ 
event-by-event over all tracks pairs, via a particle correlator $C_m$ defined
as a function of the number of accepted tracks. For purely statistical
fluctuations this observable is expected to have a null value. Once the va\-lues
are normalized to the $\langle p_T\rangle$ which is different at different
collision energies, the relative fluctuations appear to be universal at LHC 
(except at small multiplicity) and not null, as seen in Fig.~\ref{fig:fluc-norm}.
Fig.\ref{fig:fluc-norm-mc} shows comparisons with Monte Carlo at $\sqrt s$=7 TeV: PYTHIA Perugia-0
tune describes the data well except for low multiplicity, while
PHOJET significantly overestimates the observed pattern.

\section{Conclusions}
\label{sec:out}
In these proceedings we have reported an overview of ALI\-CE results in
pp collisions, primarily using the data sample collected in 2010.
Refined analyses are expected including the 2011 statistics. While various measurements
taken in pp collisions are a necessary baseline for its heavy-ions physics programme, 
a rich proton-proton physics programme has been developed by ALICE exploiting 
its detector capabilities, which are particularly relevant for soft QCD studies.

\end{document}